\pdfoutput=1

\documentclass{article}
\usepackage{spconf,amsmath,graphicx}

\usepackage{tabularx, etoolbox, booktabs}
\usepackage{multirow} 

\usepackage{rotating}
\usepackage{balance}
\usepackage{url}
\usepackage{float}  
\usepackage{placeins}  
\usepackage{stfloats}  

\usepackage{pgfplots}

\usepackage{amsmath}
\usepackage{amssymb}
\usepackage{bbm}
\usepackage{physics}
\usepackage{siunitx}
\usepackage{trfsigns}

\DeclareMathOperator*{\argmin}{argmin}

\newcommand{\vect}[1]{\ensuremath{\boldsymbol{\mathbf{#1}}}}

\usepackage{calc}
\newcounter{z}

\newcommand{\hermite}{\textcolor{lightgreen}{\mathsf H}}

\newcommand{\merlmixtures}{\textcolor{black}{WSJ0-2MIX}}
\newcommand{\reverb}{\textcolor{black}{REVERB}}
\newcommand{\smswsj}{\textcolor{black}{SMS-WSJ}}
\newcommand{\wham}{\textcolor{black}{WHAM!}}
\newcommand{\chimefive}{\textcolor{black}{CHiME5}}
\newcommand{\mird}{\textcolor{black}{MIRD}}

\newcommand{\smashEarly}[1]{$x^{\mathrm{(early)}}_{#1}$}

\usepackage{tikz}
\usepackage{amssymb}
\usetikzlibrary{calc}
\usetikzlibrary{positioning}
\usetikzlibrary{chains}
\usetikzlibrary{fit}
\usetikzlibrary{arrows}
\usetikzlibrary{decorations.pathreplacing}
\usetikzlibrary{calligraphy}
\usetikzlibrary{arrows.meta}
\usetikzlibrary{backgrounds}
\usetikzlibrary{shapes.multipart}

\tikzset{
    every text node part/.style={align=center},
    >=stealth,
    element/.style={
        draw=black!100,
        thick,
    },
    block/.style={
        element,
        rectangle,
        minimum width=4.5em,
        minimum height=2.5em,
        inner sep=0pt,
    },
    wide block/.style={
        block,
        minimum width=7.5em,
        minimum height=2.5em,
    },
    parameter/.style={
        element,
        rectangle,
        minimum width=2.5em,
        minimum height=2.5em,
        inner sep=0pt,
        text height=1.5ex,
        text depth=.25ex
    },
    random/.style={
        element,
        circle,
        minimum width=2.5em,
        minimum height=2.5em,
        inner sep=0pt,
        draw=black!100,
        text height=1.5ex,
        text depth=.25ex
    },
    observation/.style={
        random,
        double
    },
    branch/.style={
        circle,
        fill=black,
        draw=black,
        minimum size=0.2em,
        inner sep=0pt
    },
    apply/.style={
        circle,
        thick,
        draw=black!100,
        minimum size=1em,
        inner sep=0pt,
        label=center:{$\times$}
    },
    node distance=2em,
    arrow/.style={->,shorten >=0.1em},
    reverse arrow/.style={<-,shorten <=0.1em},
}

\newlength{\figurewidth}
\newlength{\figureheight}

\definecolor{yellow}{RGB}{255, 198, 0}
\definecolor{orange}{RGB}{255, 130, 0}
\definecolor{blue}{RGB}{0, 32, 91}
\definecolor{red}{RGB}{198, 53, 39}
\definecolor{magenta}{RGB}{138, 27, 97}
\definecolor{lightblue}{RGB}{0, 159, 223}
\definecolor{green}{RGB}{0, 155,119}
\definecolor{lightgreen}{RGB}{132, 189,0}
\definecolor{greenyellow}{RGB}{208, 223,0}

\usepackage[acronym, toc, nonumberlist]{glossaries}
\newacronym{BSS}{BSS}{blind source separation}
\newacronym{HMM}{HMM}{hidden Markov model}
\newacronym{SDR}{SDR}{signal to distortion ratio}
\newacronym{SIR}{SIR}{signal to interference ratio}
\newacronym{SNR}{SNR}{signal to noise ratio}
\newacronym{DC}{DC}{deep clustering}
\newacronym{DAN}{DAN}{deep attractor network}
\newacronym{PIT}{PIT}{permutation invariant training}
\newacronym{RSAN}{RSAN}{recurrent selective attention network}
\newacronym{ICA}{ICA}{independent component analysis}
\newacronym{G-cACGMM}{G-cACGMM}{Gaussian complex angular central Gaussian mixture model}
\newacronym{vMF-cACGMM}{vMF-cACGMM}{von-Mises-Fisher complex angular central Gaussian mixture model}
\newacronym{DNN}{DNN}{deep neural network}
\newacronym{TV-cGMM}{TV-cGMM}{time-variant complex Gaussian mixture model}
\newacronym{vMFMM}{vMFMM}{von-Mises-Fisher mixture model}
\newacronym{vMF}{vMF}{von-Mises-Fisher}
\newacronym{cACGMM}{cACGMM}{complex angular central Gaussian mixture model}
\newacronym{cWMM}{cWMM}{complex Watson mixture model}
\newacronym{cBMM}{cBMM}{complex Bingham mixture model}
\newacronym{EM}{EM}{expectation maximization}
\newacronym{VEM}{VEM}{variational expectation maximization}
\newacronym{MM}{MM}{majorize-minimization or minorize-maximization}
\newacronym{GEV}{GEV}{generalized eigenvalue}
\newacronym{LCMV}{LCMV}{linearly constrained minimum variance}
\newacronym{STFT}{STFT}{short time Fourier transform}
\newacronym{BLSTM}{BLSTM}{bidirectional long short term memory network}
\newacronym{FF}{FF}{feed-forward}
\newacronym{ASR}{ASR}{automatic speech recognition}
\newacronym{BAN}{BAN}{blind analytic normalization}
\newacronym{WER}{WER}{word error rate}
\newacronym{MVDR}{MVDR}{minimum variance distortionless response}
\newacronym{PESQ}{PESQ}{perceptual evaluation of speech quality}
\newacronym{PA}{PA}{permutation alignment}
\newacronym{DFT}{DFT}{discrete Fourier transformation}
\newacronym{STD}{STD}{standard deviation}
\newacronym{WSJ}{WSJ}{Wall Street Journal}
\newacronym{GMM}{GMM}{Gaussian mixture model}
\newacronym{AM}{AM}{acoustic model}
\newacronym{NMF}{NMF}{non-negative matrix factorization}
\newacronym{RIR}{RIR}{room impulse response}
\newacronym{ATF}{ATF}{acoustic transfer function}
\newacronym{RTF}{RTF}{relative transfer function}
\newacronym{EEG}{EEG}{electroencephalography}
\newacronym{DOA}{DoA}{direction of arrival}
\newacronym{TDOA}{TDoA}{time difference of arrival}
\newacronym{PDF}{PDF}{probability density function}
\newacronym{CE}{CE}{cross entropy}
\newacronym{VAD}{VAD}{voice activity detection}
\newacronym{IPD}{IPD}{intra-channel phase difference}
\newacronym{ILD}{ILD}{intra-channel level difference}
\newacronym{GCC}{GCC}{generalized cross-correlation}
\newacronym{IBM}{IBM}{ideal binary mask}
\newacronym{IRM}{IRM}{ideal ratio mask}
\newacronym{MSE}{MSE}{mean squared error}
\newacronym{WRN}{WRN}{wide residual network}

\title{SMS-WSJ: Database, performance measures, and baseline recipe\\ for multi-channel source separation and recognition}
\name{Lukas Drude, Jens Heitkaemper, Christoph Boeddeker, Reinhold Haeb-Umbach}
\address{Paderborn University, Department of Communications Engineering, Paderborn, Germany \\
\small\texttt{\{drude, heitkaemper, boeddeker, haeb\}@nt.upb.de}}
\let\OLDthebibliography\thebibliography
\renewcommand\thebibliography[1]{
  \OLDthebibliography{#1}
  \setlength{\parskip}{0.5pt}
  \setlength{\itemsep}{0.5pt plus 0.3ex}
}

\begin{document}
\ninept

\maketitle
\begin{abstract}
We present a multi-channel database of overlapping speech for training, evaluation, and detailed analysis of source separation and extraction algorithms: SMS-WSJ -- Spatialized Multi-Speaker Wall Street Journal.
It consists of artificially mixed speech taken from the WSJ database, but unlike earlier databases we consider all WSJ0+1 utterances and take care of strictly separating the speaker sets present in the training, validation and test sets.
When spatializing the data we ensure a high degree of randomness w.r.t. room size, array center and rotation, as well as speaker position.
Furthermore, this paper offers a critical assessment of recently proposed measures of source separation performance.
Alongside the  code to generate the database we provide a source separation baseline and a Kaldi recipe with competitive word error rates to provide common ground for evaluation.
\end{abstract}
\section{Introduction}
\label{sec:intro}
{\let\thefootnote\relax\footnotetext{The authors gratefully acknowledge the funding of this project by computing time provided by the Paderborn Center for Parallel Computing (PC2).}}
\Gls{BSS} aims at extracting the speech signal of individual speakers from a single- or multi-channel observation to either feed this to an \gls{ASR} system such as a speech assistant or to play it to a human listener.
In recent years, a number of different algorithms have been developed which may roughly be grouped into independent component analysis and non-negative matrix factorization-related algorithms~\cite{Hyvarinen2001ICA, Lee1999NMF, Sawada2019Review}, probabilistic spatial models~\cite{Araki2006Normalized, Mandel2010BSS, TranVu2010EM}, and neural networks~\cite{Tu2014Separation, Hershey2016DeepClustering}.

Although the different research directions have shown great progress, it is rather rare that algorithms of the different research directions are compared on a common database.
Far too often analytic approaches are compared on rather small simulated in-house databases.
It is often argued, that creating such a database is simple: a publication of the recipe to recreate it is often considered not beneficial.
In contrast, we here argue that database creation actually needs a lot of thought in order to strike a good balance between controllability and realism.
Fairly many neural network-based \gls{BSS} algorithms are evaluated on the single-channel \merlmixtures{} database published alongside~\cite{Hershey2016DeepClustering} because it provides a controlled setup with access to the source signals (compare \cite[Tbl.~1]{LeRoux2019SISDR} for an overview).
The database contains \num{20000} train mixtures of which only \num{8769} unique utterances can be used for acoustic model training.
Further, this single-channel database is noise-free and does not contain any reverberation which limits the number of algorithms which can be compared on this database as well as raises the question of how well the findings translate to more realistic scenarios.
Wang et al. provide a spatialized version of the \merlmixtures{} database~\cite{Wang2018MCDC}.
It, however, still shares the other limitations of the \merlmixtures{} database:
(a) little number of unique utterances, (b) verbalized punctuation, e.g. verbalized full-stop, (c) speakers seen during training are part of the development set.
In contrast, the proposed database removes all verbalized punctuation utterances to facilitate, e.g., sequence to sequence \gls{ASR} model training, keeps the Kaldi \gls{WSJ} split of datasets for compatibility and allows to use all remaining 33561 unique utterances for \gls{ASR} training.
The \wham{} database as well sticks to the \merlmixtures{} file lists but contains realistic background noise suitable for single-channel experiments.

The degree of realism of a database is an important factor to develop, improve, and evaluate algorithms.
The \chimefive{} data\-base \cite{Barker2018CHiME5} is fairly realistic with real recordings, frame-drop, synchronization issues, broken channels, low \gls{SNR} and spontaneous speech.
It provided a Kaldi~\cite{Povey2011Kaldi} \gls{ASR} baseline which allowed researchers to, e.g., focus on source separation while relying on the already rather elaborated \gls{ASR} baseline for evaluation.
However, since the realistic recording scenario did not allow oracles such as a clean source signal or speech images without overlap at each microphone, the realism of the database has its cost:
Metrics such as \gls{SDR} are almost impossible to obtain and the performance comparison reduces to \gls{WER} which, for many researchers, is not the measure they optimize for.

In an attempt to provide a more realistic multi-channel database than earlier mentioned databases and still have full access to all signals, this simulated database aims to find a compromise between data realism and accessibility of intermediate signals to encourage in-depth evaluation of separation algorithms:
it uses the \gls{WSJ} utterances~\cite{Paul1992WSJ} as source signals, contains simulated room impulse responses, grants access to the speech images at each microphone, and, in contrast to \cite{Seki2018Purely, Chang2019MIMO} as well to the early- and late-arriving part of the speech image.
It contrast to \cite{Seki2018Purely, Chang2019MIMO} it comes with code to extract different performance metrics and provides a rather competitive probabilistic spatial model baseline as well as a Kaldi \gls{ASR} baseline.

The entire code and well as instructions are available online:

\noindent\hspace{4em}\url{https://github.com/fgnt/sms_wsj}\hspace{4em}

\noindent The repository contains the \gls{BSS} recipe, corresponding calls to evaluation metrics, and code to train and evaluate the speech recognition baseline.
Consequently, every table in this document can be reproduced with the provided code.
The documentation and the recipes reference several external repositories and, thus, provide a convenient entry point to explore other open source contributions.

Sec.~\ref{sec:db} introduces the proposed database.
Sec.~\ref{sec:metrics} discusses performance metrics and its applicability to multi-channel recordings.
Sec.~\ref{sec:bss} and Sec.~\ref{sec:asr} introduce the source separation and speech recognition baseline.
Sec.~\ref{sec:eval} provides insights into which performance metrics are recommended and provides the evaluation results for the provided baseline system.
Finally, Sec.~\ref{sec:conclusions} concludes the proposals.

\clearpage
\section{\smswsj{} database design}
\label{sec:db}

Creating a database is a trade-off between realism and controllability.
We here opted to entirely randomize the geometric setup, simulate the room impulse responses and compose mixtures based on \gls{WSJ} utterances with compatible dataset splits.

The database consists of \num{33561}, \num{982}, and \num{1332} train, validation, and test mixtures, respectively.
The utterances were taken from the \texttt{si284}, \texttt{dev93}, and \texttt{eval92} \gls{WSJ} datasets\footnote{Naming according to the Kaldi WSJ recipe.} and downsampled to \SI{8}{kHz}.
To facilitate acoustic model training it is ensured that the sets contain as many unique utterances as possible (Each unique utterance is repeated equally often.): the sets contain \num{33561}, \num{491}, and \num{333} unique utterances.
In contrast, the \merlmixtures{}~\cite{Hershey2016DeepClustering} and its spatialized counterpart~\cite{Wang2018MCDC} have \num{20000} mixtures but only \num{8769} unique utterances.
Further, we excluded utterances with verbalized punctuation to facilitate training of, e.g., CTC or sequence-to-sequence acoustic models and to avoid using further filtering.

The length of each mixture is determined by the longest utterance.
The shorter utterance is zero-padded with a random uniform offset.
Fig.~\ref{fig:hist_overlapp} shows the relative overlap of the utterances.
The overlap was measured by first identifying the actual beginning and ending of each utterance by analyzing the silence alignments produced by an acoustic model operating on oracle signals.

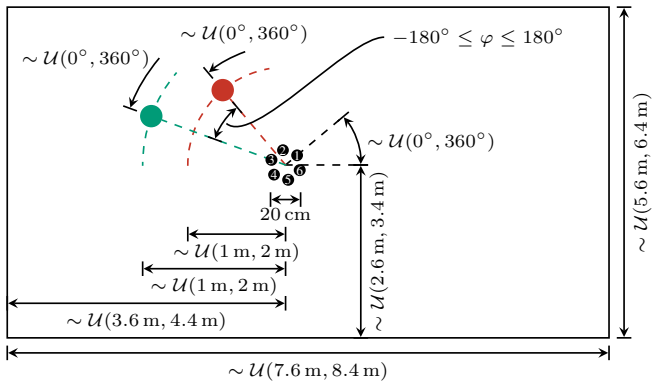
\begin{figure}[b]
\vspace{-0.5em}
\centering
\begin{tikzpicture}[
    x=0.1mm,
    y=0.1mm,
    line width=0.07em,
    shorten >=-0.035em,  
    shorten <=-0.035em,
]
\sisetup{parse-numbers=false}

\tikzstyle{every node}=[font=\scriptsize]

\coordinate (roomSouthWest) at (0, 0);
\draw (0, 0) -- (800, 0) -- (800, 440) -| cycle;
\draw [|<->|, xshift=2mm] (800, 0) -- node [below, sloped] { $\sim\mathcal{U}(\SI{5.6}{m}, \SI{6.4}{m})$} (800, 440);
\draw [|<->|, yshift=-2mm] (0, 0) -- node [below] {$\sim\mathcal{U}(\SI{7.6}{m}, \SI{8.4}{m})$} (800, 0);

\coordinate (array) at (370, 230);
\draw [|<->|] ($(array) + (100, 0)$) arc(0:40:100) node [near end, right, xshift=1mm, yshift=-2mm] {$\sim\mathcal{U}(\SI{0}{\degree}, \SI{360}{\degree})$} coordinate (helper);

\draw [dashed] (array) -- ($(array) + (100, 0)$);
\draw [dashed] (array) -- (helper);

\coordinate (source) at (array);
\node [circle, fill=red] (source1) at ($(source) + (130:130)$) {};
\node [circle, fill=green] (source2) at ($(source) + (160:190)$) {};

\draw [dashed, red] (source1) arc(130:100:130);
\draw [dashed, red] (source1) arc(130:180:130);
\draw [dashed, green] (source2) arc(160:140:190);
\draw [dashed, green] (source2) arc(160:180:190);

\draw [dashed, red] (source) -- (source1);
\draw [dashed, green] (source) -- (source2);

\draw [->|] ($(source) + (110:160)$) node [above] {$\sim\mathcal{U}(\SI{0}{\degree}, \SI{360}{\degree})$} arc(110:130:160);
\draw [->|] ($(source) + (140:220)$) node [left] {$\sim\mathcal{U}(\SI{0}{\degree}, \SI{360}{\degree})$} arc(140:160:220);

\draw [|<->|] ($(source)!8/40!(source|-roomSouthWest) + (-20, 0)$) -- node [below] {$\SI{20}{cm}$} +(40, 0);

\draw [|<->|] ($(source)!16/40!(source|-roomSouthWest)$) -- node [below] {$\sim \mathcal U(\SI{1}{m}, \SI{2}{m})$} +(-130, 0);
\draw [|<->|] ($(source)!24/40!(source|-roomSouthWest)$) -- node [below] {$\sim \mathcal U(\SI{1}{m}, \SI{2}{m})$} +(-190, 0);

\draw [|<->|] ($(array) + (130:100)$) arc(130:160:100);
\draw ($(array) + (145:95)$) to[out=-40, looseness=0.5, in=180] (500, 400) node [right] {$\SI{-180}{\degree}\leq \varphi \leq \SI{180}{\degree}$};

\foreach \x in {1, ..., 6}{
    \node [circle, fill=black, minimum width=1.5mm, inner sep=0.01em, text=white] at ($(array) + (60*\x-60+40:20)$) {\tiny \x};
}


\draw [|<->|] ($(source)!32/40!(source|-roomSouthWest)$) -- node [below] {$\sim\mathcal{U}(\SI{3.6}{m}, \SI{4.4}{m})$} ($(source-|roomSouthWest)!32/40!(roomSouthWest)$);

\draw [|<->|] (470, 0) -- node [below, sloped] {$\sim\mathcal{U}(\SI{2.6}{m}, \SI{3.4}{m})$} (470, 230);
\end{tikzpicture}
\vspace{-2em}
\caption{%
Geometry of the SMS-WSJ database.
Each aspect of the geometry is uniformly sampled from the given range.
The array is randomly rotated along all three geometric axes.
Only the $z$-axis rotation is shown in this figure.
}
\label{fig:wsjbss}
\end{figure}

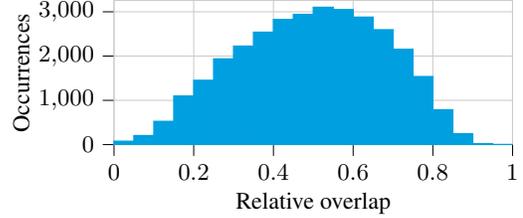
\begin{figure}[t]
\setlength\figureheight{3.5cm}
\setlength\figurewidth{0.8\columnwidth}
\centering
\begin{tikzpicture}

\begin{axis}[
height=\figureheight,
width=\figurewidth,
tick align=outside,
tick pos=left,
xmajorgrids,
x grid style={white!80.0!black},
ymajorgrids,
y grid style={white!80.0!black},
axis line style={white!80.0!black},
xmin=0, xmax=1,
xlabel={Relative overlap},
ylabel={Occurrences},
xtick style={color=black},
ytick style={color=black},
ymin=0, ymax=3257.1,
]
\tikzstyle{myStyle}=[fill=lightblue, draw=lightblue,draw opacity=1, line width=0.5pt]

\draw[myStyle] (axis cs:0,0) rectangle (axis cs:0.05,76);
\draw[myStyle] (axis cs:0.05,0) rectangle (axis cs:0.1,204);
\draw[myStyle] (axis cs:0.1,0) rectangle (axis cs:0.15,525);
\draw[myStyle] (axis cs:0.15,0) rectangle (axis cs:0.2,1097);
\draw[myStyle] (axis cs:0.2,0) rectangle (axis cs:0.25,1452);
\draw[myStyle] (axis cs:0.25,0) rectangle (axis cs:0.3,1937);
\draw[myStyle] (axis cs:0.3,0) rectangle (axis cs:0.35,2224);
\draw[myStyle] (axis cs:0.35,0) rectangle (axis cs:0.4,2539);
\draw[myStyle] (axis cs:0.4,0) rectangle (axis cs:0.45,2826);
\draw[myStyle] (axis cs:0.45,0) rectangle (axis cs:0.5,2941);
\draw[myStyle] (axis cs:0.5,0) rectangle (axis cs:0.55,3102);
\draw[myStyle] (axis cs:0.55,0) rectangle (axis cs:0.6,3050);
\draw[myStyle] (axis cs:0.6,0) rectangle (axis cs:0.65,2879);
\draw[myStyle] (axis cs:0.65,0) rectangle (axis cs:0.7,2595);
\draw[myStyle] (axis cs:0.7,0) rectangle (axis cs:0.75,2152);
\draw[myStyle] (axis cs:0.75,0) rectangle (axis cs:0.8,1537);
\draw[myStyle] (axis cs:0.8,0) rectangle (axis cs:0.85,786);
\draw[myStyle] (axis cs:0.85,0) rectangle (axis cs:0.9,246);
\draw[myStyle] (axis cs:0.9,0) rectangle (axis cs:0.95,21);
\draw[myStyle] (axis cs:0.95,0) rectangle (axis cs:1,1);
\end{axis}

\end{tikzpicture}
\vspace{-1em}
\caption{%
Histogram of relative overlap between two active speakers.
}
\vspace{-1em}
\label{fig:hist_overlapp}
\end{figure}

The geometric setup is randomly sampled, such that the room size, the array center, and the array rotation is random.
The distance of each source around the array center is then randomly sampled from $\mathcal U(\SI{1}{m}, \SI{2}{m})$.
Subsequently, the azimuth angle of each source around the center is uniformly sampled without enforcing any kind of minimum angular distance, i.e., two sources can potentially be behind each other as in the \mird{} database~\cite{Hadad2014Database}.
The proposed database allows to analyze the separation performance as a function of angular distance.
The sensor array itself is simulated as a circular array with radius $\SI{10}{cm}$.
It is taken care that odd behavior in the \gls{RIR} simulation due to accidental symmetries in the geometric setup is avoided, e.g., the connecting line between a source and a sensor orthogonal with a wall has probability zero by avoiding discrete positions and an additional slight random tilt of the circular array.
Fig.~\ref{fig:wsjbss} illustrates the geometric setup.

The \glspl{RIR} were generated using the image method~\cite{Allen1979Image} with the implementation from~\cite{Habets2006Image}\footnote{We here provide a thin Python wrapper for \cite{Habets2006Image}.} with a random sound decay time (T60) sampled from $\mathcal U(\SI{200}{ms}, \SI{500}{ms})$.
It is ensured, that the offset introduced due to time of flight is compensated for all channels at once to avoid an unrealistic manipulation of the spatial characteristics.

The mixing process in time domain is simulated as follows:
\begin{align}
\vect y_\ell = \begin{bmatrix}y_{\ell, d=0}\\\vdots\\y_{\ell, d=D}\end{bmatrix}
= \sum_k \vect x_{k, \ell} + \vect n_\ell
= \sum_k \vect h_\ell \ast s_{k, \ell} + \vect n_\ell,
\end{align}
where $\vect y_\ell$ is the observed signal vector at the $D$ microphones for the sample index $\ell$, $\vect x_{k, \ell}$ is the source image of source $k$, $s_{k, \ell}$ is the source signal at its origin, and $\vect n_\ell$ is artificially generated white sensor noise with a rather low \gls{SNR} randomly sampled from $\mathcal U(\SI{20}{dB}, \SI{30}{dB})$.
The $*$ operator is a convolution.
We opted for white sensor noise because a spatially realistic simulation of background noise is still an unsolved problem, e.g., although the \wham{} database provides real background noise recordings for a fixed sensor array~\cite{Wichern2019WHAM}, however, the background noise can not be trivially used for other geometries and would not match the simulated \glspl{RIR} of the speech sources.

\begin{figure}[b]
\vspace{-1.5em}
\setlength\figureheight{4cm}
\setlength\figurewidth{1\columnwidth}
\centering
\input{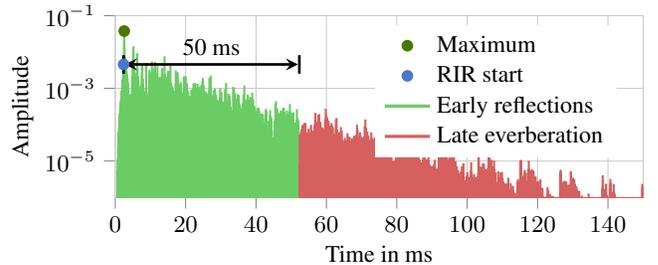}
\vspace{-1em}
\caption{%
Semi-logarithmic plot of an example \gls{RIR}.
Differentiating between the early part and late part of the \gls{RIR} allows to better analyze the dereverberation properties of an algorithm or even train neural networks to predict the less reverberant parts of speech.
}
\label{fig:rir}
\end{figure}

Thanks to the simulated nature of this database, the speech images $\vect x$ at each microphone can further be decomposed in an early-arriving part of the signal and a late-arriving part of the signal:
\begin{align}
\vect x_{k, \ell}
= \vect x^{(\mathrm{early})}_{k, \ell} + \vect x^{(\mathrm{late})}_{k, \ell}
=
\vect h^{(\mathrm{early})}_\ell \ast s_{k, \ell} + \vect h^{(\mathrm{late})}_\ell \ast s_{k, \ell}.
\end{align}
The \gls{RIR} start sample was determined by finding the first sample which is larger than the maximum divided by ten.
To consider all microphones, the \gls{RIR} start was selected as the smallest value across all microphones. This value is used to remove the propagation delay in $\vect x$.
The end of the early part of the \gls{RIR} was set to be \SI{50}{ms} after the start sample inspired by the definition in the \reverb{} database~\cite{Kinoshita2013ReverbChallenge} and the precedence effect~\cite{Litovsky1999Precedence}.
This allows to evaluate dereverberation capabilities of the developed systems as well as training of, e.g., a neural network to predict the early-arriving part of a speech signal~\cite{Heymann2017Neural}.
Fig.~\ref{fig:rir} illustrates an example \gls{RIR}.


\clearpage
\section{Evaluation metrics}
\label{sec:metrics}

To allow consistent comparison of different algorithms we here discuss and recommend a selection of performance metrics applicable in a multi-speaker multi-channel far-field scenario.
For the sake of simplicity, the here presented equations for the metrics are only defined for a single speaker and ignore the permutation problem.

First of all, it is worth noting, that selecting the right performance metric is an ongoing debate~\cite{Mandel2010Evaluating, LeRoux2019SISDR}.
Therefore, we first discuss different \gls{SDR} variants.
In its most restrictive form \gls{SDR} can be defined as the ratio of the power of the signal of interest (here the speech source signal, not the speech image) and the power of that part of the signal of interest, which cannot be explained by the prediction~\cite[Eq.~1]{LeRoux2019SISDR}, \cite[Sec.~II.]{Vincent2006Performance}:
\begin{align}
\frac{\sum_\ell | s_{k, \ell} |^2}{\sum_\ell | s_{k, \ell} - \hat s_{k, \ell} |^2}.
\end{align}
This metric is appropriate, when it can be assured that the oracle signal and the estimate have matching scaling.
This can be the case for, e.g., single-channel applications where an anechoic signal is corrupted by additive noise.
A delay, a short \gls{RIR} or a gain factor would be penalized heavily.
A somewhat less strict \gls{SDR} definition allows an arbitrary gain mismatch by scaling the prediction to match the scale of the signal of interest.
To stay consistent with~\cite[Eq.~2 and 3]{LeRoux2019SISDR}, we call this SI-SDR here:
\begin{align}
\frac{\sum_\ell | s_k |^2}{\sum_\ell | s_{k, \ell} - \beta \hat s_{k, \ell} |^2}
~&\mathrm{for}~
\beta ~\mathrm{such~that}~  s_{k, \ell} \perp (s_{k, \ell} - \beta \hat s_{k, \ell}) \\
=
\frac{\sum_\ell | \alpha s_{k, \ell} |^2}{\sum_\ell | \alpha s_{k, \ell} - \hat s_{k, \ell} |^2}
~&\mathrm{for}~
\alpha = \argmin_{\alpha}  | \alpha s_{k, \ell} - \hat s_{k, \ell} |^2
.
\end{align}
This SI-SDR metric is applicable, when a constant gain is expected, e.g. when a regression model uses a variance normalization.
BSS-Eval SDR, as defined in~\cite[Sec.~III.B.]{Vincent2006Performance} allows filtering with an arbitrary impulse response up to a maximum length $\tau_{\mathrm{max}} = 512$:
\begin{align}
\frac{\sum_\ell | \alpha_\ell \!*\! s_{k, \ell} |^2}{\sum_\ell | \alpha_\ell \!*\! s_{k, \ell} \!-\! \hat s_{k, \ell} |^2}
~\mathrm{for}~
\alpha_\ell = \argmin_{\alpha_\ell} \sum_\ell | \alpha_\ell \!*\! s_{k, \ell} \!-\! \hat s_{k, \ell} |^2
\!\!.\!\!
\end{align}
Consequently, a gain, slight time delay as well as equalization effects do not harm the actual metric.
This allows, e.g., a low pass filter, but counts late reverberation as artifacts~\cite{Mandel2010Evaluating}.
This is criticized in~\cite[Sec.~2.4.]{LeRoux2019SISDR} since an enhancement system may exploit the invariance wrt. such an impulse response.
However, as will be discussed in Sec.~\ref{sec:eval}, a certain linear filtering effect, such as switching the reference microphone in a compact array, should not influence the performance metric.

The \emph{invasive SDR} metric assumes a linear enhancement, e.g. channel selection and mask multiplication, beamforming, or dereverberation.
To compute the score the enhancement $O_k\{\}$ with parameters estimated on the observation aims to extract source $k$ and is applied to individual signal components independently:
\begin{align}
  \frac{\sum_\ell \left| O_k\left\{\vect x_{k, \ell}\right\}\right|^2}{\sum_\ell (\sum_{\tilde{k} \ne k} \left| O_k\left\{\vect x_{\tilde{k}, \ell}\right\}\right|^2 + \left| O_k\left\{\vect n_{\tilde{k}, \ell}\right\} \right|^2)}
\end{align}
Variants of invasive SDR are extensively used in beamforming literature.
Its main advantage is the avoidance of any estimation or projection as in BSS-Eval SDR and its natural extension to multiple channels.
The database at hand supports this metric due to access to the speech images.

Both perceptual evaluation of speech quality (PESQ)~\cite{Rix2001PESQ}, which was originally developed for telephone channel evaluation and short time objective intelligibility (STOI)~\cite{Taal2011STOI} aim at measuring the perceptual quality of speech and, as such, provide a complementary metric to the \gls{SDR} metrics.
A system which exploits deficiencies of an \gls{SDR} metric, e.g., a system eliminating entire frequencies and produces overly optimistic \gls{SDR} values~\cite[Fig.~2]{LeRoux2019SISDR} is likely to perform poorly in terms of PESQ and STOI.

Finally, evaluating the performance of a speech enhancement or source separation front-end with an acoustic model in terms of \glspl{WER} comes with its own set of advantages and disadvantages.
First of all, many minor improvements of the speech signal by the front-end may be eaten up by a strong acoustic model: stronger acoustic models favor front-ends which exploit cues which the acoustic model does not have access to.
Nevertheless, \gls{WER} is one of the hardest metrics to cheat: a front-end which results in excellent \gls{WER} is rarely just exploiting some specifics of the metric.

\vspace{-0.5em}
\section{Source separation baseline recipe}
\label{sec:bss}
\vspace{-0.5em}
The utility of a database stands and falls with the availability of a reasonable baseline.
Therefore, we provide a \gls{BSS} recipe consisting of a spatial clustering followed by a beamforming operation.

The spatial clustering model is a \gls{cACGMM} operating on all microphone channels~\cite{Ito2016cACGMM}.
It is defined by the following marginal distribution with a time-dependent mixture weight~\cite{Ito2013permutation}:
\vspace{-0.75em}
\begin{align}
p(\vect y_{t,f}) = \sum_{k=1}^{K+1} \pi_{k, t} \frac{(D-1)!}{2\pi^{D}\det \vect B_{k,f}} \frac{1}{(\vect y_{t,f}^{\mathsf H} \vect B_{k,f}^{-1} \vect y_{t,f}^{\vphantom{\hermite}})^{D}}.
\end{align}
\vspace{-0.8em}

The mixture model is initialized randomly with $K+1$ classes (an extra class for the noise estimate) such that the class affiliation posterior of each time frequency bin is independently sampled from a uniform $(K+1)$-dimensional Dirichlet distribution.
The parameters of the model are then estimated using the \gls{EM} algorithm
with an inline permutation alignment step~\cite{Sawada2007Permutation} for the time-dependent mixture weight.
After convergence, the \gls{cACGMM} yields class affiliation posteriors which can then,
be used for a mask-based beamforming.
Here, we use Souden's formulation of the \gls{MVDR} beamformer~\cite{Souden2010MVDR} with a reference channel selection based on expected output SNR~\cite[Sec.~4]{Erdogan2016MVDR}.
The speech and distortion covariance matrices are estimated as the weighted mean of outer products of the observation vector $\vect y_{t, f}$ using masks obtained from the mixture model.
The mask for the distortion matrix is obtained by summing up all masks not belonging to the target speaker.

\vspace{-0.5em}
\section{Speech recognition baseline recipe}
\label{sec:asr}
\vspace{-0.5em}
The baseline acoustic model is a factorized time-delayed neural network (TDNN-F) based on the WSJ Kaldi recipe~\cite{Povey2011Kaldi}.
For bootstrapping, a GMM-HMM system is trained first.
Then, due to the nature of this database, the alignments are extracted on the early-arriving speech image $\smash{x_{k=0, \ell, d}^{(\mathrm{early})}}$.
This has the advantage, that the alignments have the appropriate time shift for the speech images, i.e., the propagation delay and random start and stop times are already accounted for.
These alignments are then used to train an acoustic model consisting of 8 TDNN-F layers.
The reverberant noisy images $x_{k, \ell, d} + n_{\ell, d}$ for multiple channels and both speakers are chosen to ensures that the acoustic model saw as much variability during training as possible, while not training on separation results.
The final word sequence is obtained by decoding the state posteriors with a default WSJ tri-gram Kaldi language model without an additional n-best rescoring.

\clearpage
\begin{table*}[t!]
\caption{%
Comparison of the specifics of different metrics.
The metrics are extracted and averaged for the both speaker with varying oracle predictions (rows) and varying reference signals (columns).
The time index $\ell$ and speaker index $k$ are dropped for clarity.
}
\label{tbl:eval_metrics}
\centering
\renewcommand*{\arraystretch}{1.25}
\footnotesize

\begin{tabular}{
ll
p{0em}
S[table-auto-round, table-format=-2.1]
S[table-auto-round, table-format=-2.1]
S[table-auto-round, table-format=-2.1]
p{0em}
S[table-auto-round, table-format=-2.1]
S[table-auto-round, table-format=-2.1]
S[table-auto-round, table-format=-2.1]
p{0em}
S[table-auto-round, table-format=2.1]
S[table-auto-round, table-format=2.1]
S[table-auto-round, table-format=2.1]
S[table-auto-round, table-format=2.1]
}
\toprule
& &&
\multicolumn{3}{c}{Reference for SI-SDR / \si{dB}} &
\multicolumn{5}{c}{\!\!Reference for BSS-Eval SDR / \si{dB}\!\!} &
\multicolumn{4}{c}{Training data for WER / \si{\percent}} \\
\cmidrule{4-6}
\cmidrule{8-10}
\cmidrule{12-15}
& &&
{$\smash{s}$} &  {\smashEarly{d=0}} &   {$\smash{x_{d=0}}$} &&
{$\smash{s}$} &  {\smashEarly{d=0}} &   {$\smash{x_{d=0}}$} &&
{$\smash{s}$} &  {\smashEarly{d}} &   {$\smash{x_{d}}$} & {$\smash{x_{d} + n_{d}}$} \\
\midrule
\multirow{7}{*}[-3ex]{\rotatebox{90}{Predict candidate}} & {$s$}
                       &&      {inf} & -18.073245 & -18.447564 && 275.262092 &  -1.981068 &  -2.695393 && 5.34 & 7.19 & 5.87 & 6.79 \\
\cmidrule{2-15}
& {\smashEarly{d=0}}   && -18.073245 &      {inf} &  11.830712 && 54.418243 & 266.278354 &  15.280209 && 13.98 & 6.32 & 6.74 & 7.34 \\
& {\smashEarly{d=1}}   && -18.304128 &  -0.248747 &  -1.111580 && 54.693485 &  10.010003 &   7.733374 && 14.13 & 6.32 & 6.77 & 7.33 \\
\cmidrule{2-15}
& $x_{d=0}$            && -18.447564 &  11.830712 &      {inf} && 14.928220 &  15.755578 & 266.510031 && 46.13 & 19.37 & 7.78 & 8.73 \\
& $x_{d=1}$            && -18.633673 &  -1.110547 &  -0.348366 && 14.953767 &   8.537105 &   8.361173 && 46.25 & 19.29 & 7.71 & 8.45 \\
\cmidrule{2-15}
& $x_{d=0} + n_{d=0}$  && -18.495680 &  11.023076 &  21.886414 && 13.463826 &  14.141180 &  21.909572 && 51.95 & 25.96 & 12.81 & 8.95 \\
& $x_{d=1} + n_{d=1}$  && -18.665072 &  -1.185733 &  -0.431150 && 13.480833 &   8.114920 &   7.946574 && 52.10
 & 25.90 & 12.68 & 9.02 \\
\bottomrule

\end{tabular}

\vspace{-1em}
\end{table*}

\section{Evaluation}
\label{sec:eval}
\vspace{-0.2em}
The evaluation section is split in two parts.
First, we investigate how different \gls{SDR} variants react to the reverberant far-field scenario.
Then, we compare the proposed baseline with different oracles.

Tbl.~\ref{tbl:eval_metrics} provides detailed performance metrics in case a certain oracle is used as the prediction signal (rows) when a certain oracle is used as the reference (columns).
We will now dissect the table step by step to shed light on which reference to use and which performance metrics is favorable.

First of all, we notice the healthy property that both SI-SDR as well as BSS-Eval SDR have extremely high values when the prediction and the reference coincide.
We can now investigate, how the different performance metrics change, when we switch the channel of the oracle.
Since we operate with a rather compact array, the difference between channel 0 and channel 1 is inaudible.
Consequently, changing the channel of the predicted image from $x_0$ to $x_1$ does not change the word error rate, no matter on what the acoustic model was trained.
However, the SI-SDR changes dramatically from infinity to \SI[round-precision = 1,]{-0.348366}{dB} when the oracle prediction system predict the speech image at sensor 1 instead of sensor 0 and the reference is sensor 0.
Thus, we need to find a reference signal and a performance metric, which changes only little, when a system perfectly predicts the speech image, just not at the reference sensor.
We quickly notice, that BSS-Eval SDR with the source signal $s$ as a reference has this favorable property: both early-arriving speech images have around \SI{54}{dB}, both speech images have \SI{15}{dB} and both noisy observations (without interfering speaker) have \SI[round-precision = 1]{13.5}{dB}.
The further we deviate from the source signal $s$, the lower the metric is.
In contrast, SI-SDR has around \SI{-18}{dB} for almost all oracle predictions when using the source signal $s$ as a reference.
This is due to the fact, that SI-SDR does not allow deviations explained by a short FIR filter.
BSS-Eval SDR allows a short FIR filter with a maximum delay of 512 samples (here \SI{64}{ms}).
It is worth noting, that the actual time of flight is already compensated by the database design.
The behavior of SI-SDR can thus not be attributed to the time of flight.

In general, we notice that the \glspl{WER} are rather stable with respect to which channel the oracle prediction system produces.
Further, we realize that the best \glspl{WER} for a specific input are obtained with matched training.
We may conclude, that the training data should not be cleaner than the test data, e.g., we should not train on $s$ when we expect a \gls{BSS} prediction to be closer to a noisy image.
In most practical cases, when the acoustic model is not retrained on the enhanced data, it is advisable to use multi-condition training, e.g., expose the acoustic model to as much variability as possible.

In conclusion of the metric discussion, we first of all recommend to use more than one metric including \gls{WER}: this dramatically reduces the likelihood, that a system exploits specifics of one particular metric.
Further, we recommend to use BSS-Eval SDR with the source signal $s$ over SI-SDR to evaluate far-field scenarios.

\begin{table}[b]
\vspace{-2em}
\caption{%
Test results of the baseline recipe.
Oracles are denoted in gray.
All metrics are averaged across all utterances and speakers.
}
\label{tbl:baseline}
\renewcommand*{\arraystretch}{1.1}
\centering
\footnotesize
\begin{tabular}{
l
S[table-auto-round, table-format=-2.1]
S[table-auto-round, table-format=-2.1]
S[table-auto-round, table-format=1.2]
S[table-auto-round, table-format=1.2]
S[table-auto-round, table-format=2.2]
}
\toprule
System & \multicolumn{2}{c}{SDR / \si{dB}} & {PESQ} & {STOI} & {WER} \\
\cmidrule{2-3}
& {\!\!\!\!BSS-Eval\!\!\!\!} & {\!\!\!\!Invasive\!\!\!\!} & & & {/ \si{\percent}} \\
\midrule
$\smash{y_{l, d=0}}$ & -0.39725325754478846 & -0.03578441058533363 & 1.4958036786786786 & 0.6587315736709655 & 79.03\\
\midrule
\smash{MM, Masking} & 9.546185730236255 & 13.88230642553854 & 1.8267023273273277 & 0.7799839502265367 & 39.0 \\
\smash{MM, MVDR} & 12.321637151158157 & 15.71628928213399 & 2.0633607357357358 & 0.8195450702496488 & 18.7 \\
\midrule
\color{gray}\smash{IRM, MVDR} & \color{gray}12.451142997544075 & \color{gray}15.657375371659205 & \color{gray}2.020460960960961 & \color{gray}0.8227555044149155 & \color{gray}17.19 \\
\color{gray}\smash{IBM, MVDR} & \color{gray}12.876207284443112 & \color{gray}16.918707934098418 & \color{gray}2.0592935435435438 & \color{gray}0.8294738196402653 & \color{gray}14.5 \\
\color{gray}\raisebox{0.1ex}{$x_{k, \ell, d=0}$} & \color{gray}14.928219751405225 & \color{gray}{~~~~~~n/a} & \color{gray}2.0525439189189187 & \color{gray}0.8313851174441846 & \color{gray}8.73 \\
\color{gray}{\raisebox{-0.1ex}{\smash{\smashEarly{k, \ell, d=0}}}} & \color{gray}54.41824292905 & \color{gray}{~~~~~~n/a} & \color{gray}2.3499320570570568 & \color{gray}0.8623779952729939 & \color{gray}7.34 \\
\bottomrule

\end{tabular}
\end{table}

Now, we evaluate the baseline recipe described in Sec.~\ref{sec:bss} and \ref{sec:asr}.
First, \gls{STFT} signals for the \gls{BSS} algorithm were extracted using a Hann window, a window size of 512, a \gls{DFT} size of 512, and a shift of 128 for the \SI{8}{kHz} signals.
After applying the beamforming operation, the signals are transformed back to the discrete time domain before the 40 variance normalized MFCC features were extracted.
All results are presented on the test set with the language model weight selected on the validation set.
Tbl.~\ref{tbl:baseline} shows the metrics without any \gls{BSS} algorithm in the first row.
The second and third row shows the results for the baseline \gls{cACGMM} with masking on channel 0 and with beamforming, respectively.
The beamforming results, in particular in terms of \gls{WER}, are clearly better than the masking results.
The oracle results (in gray) indicate that there is sufficient room for (a) better source separation and (b) masks more appropriate for masking and/or beamforming.

\vspace{-0.6em}
\section{Conclusions}
\label{sec:conclusions}
\vspace{-0.3em}
In this contribution we introduced a simulated database to allow comparison of \gls{BSS} algorithms in a far-field scenario.
We provide a \gls{BSS} baseline and a competitive acoustic model to facilitate detailed comparisons of \gls{BSS} algorithms with a given acoustic model.
Further we discussed and analyzed different performance metrics and provide concrete recommendations: use multiple complementary metrics including \gls{WER} and prefer BSS-Eval SDR over SI-SDR in a far-field scenario.

\clearpage
\balance
\bibliographystyle{IEEEbib}
\bibliography{strings,refs}

\end{document}